\documentclass[aps,twocolumn,amsmath,amssymb,floatfix,fleqn]{revtex4}
\usepackage{lineno,hyperref}
\usepackage{graphicx}
\usepackage{bm}
\usepackage[german,american]{babel}
\usepackage{bigints}
\usepackage{mathtools, nccmath}
\usepackage{tabularx}
\usepackage{color}
\newcommand{\fig}[1]{Fig.\ \ref{#1}}
\newcommand{\eqn}[1]{Eq.\ \eqref{#1}}

\usepackage{accents}
\newcommand*{\dt}[1]{%
  \accentset{\mbox{\large\bfseries .}}{#1}}

\makeatletter
\DeclareRobustCommand{\iscircle}{\mathord{\mathpalette\is@circle\relax}}
\newcommand\is@circle[2]{%
  \begingroup
  \sbox\z@{\raisebox{\depth}{$\m@th#1\bigcirc$}}%
  \sbox\tw@{$#1\square$}%
  \resizebox{!}{\ht\tw@}{\usebox{\z@}}%
  \endgroup
}
\makeatother

\begin{document}

\title{
Extracting Rates and Activation Free Energies of Martensitic Transitions Using
Nanomechanical Force Statistics: Theory, Models, and Analysis
}
\author{Arijit Maitra}
\email{arijit.maitra@bmu.edu.in}
\affiliation{Department of Applied Sciences, School of Engineering and Technology, BML Munjal University, NH 8, 67 KM Milestone, Gurugram, Haryana 122413, India }

\author{M.\ P.\ Gururajan}
\affiliation{Department of Metallurgical Engineering and Materials Science, Indian Institute of Technology Bombay,
Mumbai, Maharashtra 400076, India }

\date{\today}

\begin{abstract}

\noindent
Nanomechanical responses (force-time profiles) of crystal lattices 
under deformation exhibit random \emph{critical} jumps, reflecting 
the underlying structural transition processes.
Despite extensive data collection, interpreting dynamic critical responses 
and their underlying mechanisms remains a significant challenge.
This study explores a microscopic theoretical approach
to analyse critical force fluctuations 
in martensitic transitions. 
Extensive sampling of the critical forces was performed 
using nonequilibrium molecular dynamics simulations 
of an atomic model of single-crystalline titanium nickel.
We demonstrate that a framework of nonequilibrium statistical mechanics 
offers a principled explanation of the relationship 
between strain rate and the critical force distribution 
as well as its mean.
The martensitic transition is represented on a free energy landscape, taking into account the thermally activated evolution of atomic arrangements over a barrier during its time-dependent deformation.
The framework enables consistent inference of the relevant fundamental properties (e.g., intrinsic rate, activation free energy) that define the rate process of structural transition. 
The study demonstrates how the statistical characterisation of nanomechanical response-stimulus patterns can offer microscopic insights into the deformation behaviours of crystalline materials.

\end{abstract}

\maketitle

\section{Introduction}
\sloppy 

The development of \emph{in situ} nanomechanical techniques \cite{Dehm2018,Kiener2024} 
has widened the scope 
for evaluating materials, 
enabling the exploration of
their mechanical and kinetic properties at the nano and submicron scale.
These techniques include nano/micro compression \cite{Dimiduk2005,FRICK2008}, 
scanning probe microscopy \cite{Binning,Zhong2024},
microelectromechanical system (MEMS) \cite{Bhowmick2019,Zhu2005} and
nanoindentation \cite{Oliver1992,Schuh2006,Kaushik2022}
for precise manipulation of crystal lattices.
This is achieved by directly 
applying forces (loads) or displacements to the specimen
and measuring the evolution of its response as the underlying crystal lattice progressively deforms. 
Metallic lattices under constant loading rates (the focus of this study) exhibit critical stochastic discontinuities in their nanoscale force-time responses \cite{uchic2004}, reflecting the underlying mechanism of deformation.
These mechanisms involve 
changes in chemical bond orientations 
or unit-cell structures along specific reaction coordinates, 
causing defects (e.g., stacking faults, twins, dislocations, etc.) \cite{Bhadeshia_geometry} 
to form in the lattice structure. 
Microscopic studies are crucial for understanding the observed stochasticity, 
its physical relevance and association with specific 
deformation mechanisms \cite{Schuh2004,Morris2011,Friedman2012,Pattamatta2014}. 
Currently, such studies are gaining increasing importance. 
We address this here through an
analytical approach
based on non-equilibrium statistical physics,
which allows modelling the statistically 
discrete nature of the force-response behaviour 
by considering the energetic 
and kinetic properties governing the microscopic structural transitions. 

Since microscopic transition dynamics are intrinsically stochastic, 
material geometries configured at very small scales are susceptible 
to manifest phase transition-linked uncertainties that 
eventually affect output performance when loaded. 
We examine theoretically such uncertainties in single crystalline titanium nickel \cite{Otsuka2005} 
under various rates of pseudoelastic tensile deformation. 
The critical discontinuities were extensively 
sampled through atomistic molecular dynamics simulations,
which showed that the force-drop events in the 
responses coincided with a martensitic phase transition 
in the lattice, triggering the formation of \emph{twin} defects 
\cite{Christian1995,Beyerlein2014,Uttam2020}.
We analyse and model the critical patterns of force responses
underlying the twinning phenomenon, 
which holds significant relevance in structural and functional materials.

In pseudoelastic titanium nickel, a time-dependent mechanical bias
imposed on it,
can drive austenite-to-martensite \emph{phase transitions} (isothermal)
and locally reshape the lattice unit cells
 through bond length and orientation changes,
producing characteristic twinned domains in the 
microstructure.
These interactions 
culminate in driving shear displacements of the atomic planes, 
forming specific planes with mirror symmetry, known as twin planes.
Dynamically regulating twinning's remarkable reversibility property
through controlled mechanical bias and thermal protocols,
the symmetry of crystal structures can be manipulated to
transduce energy into useful work \cite{Bhattacharya2004,Bhattacharya2005}.
The profound interest in converting these materials into nano and micro machines,
including applications, e.g., actuation (force or motion generation), force-damping,
and other 'smart' end-uses \cite{MohdJani2014,McCracken2020}
for the robotics, aerospace and biomedical industries,
is rooted in this principle.

Solid-solid phase transition-mediated twinning mechanisms 
have been the subject of many insightful theoretical and 
computational studies \cite{Chowdhury2017,Niitsu2020,Yan2016}.
First-principles calculations \cite{Kumar2020}, density functional theory (DFT) 
\cite{Ogata2005,Hatcherprb2009,GudaVishnu2010,Zarkevich2014}, molecular dynamics simulations 
\cite{Li2020,Tang2018} 
and thermodynamic modelling \cite{Muller2001,Falk1980,Falk1983} approaches have provided
critical insights into energy landscapes, phase stability, 
and pathways for martensitic transformations in various metals, including titanium nickel.
Recent findings have highlighted the dynamic nature of martensitic phase transition and its modulation
by force 
\cite{Maitra2022}. 
As our interest lies in inference-based property calculation from measured 
distributions,
including experimental data, here we employ principles focused on resolving 
the dynamical information encoded in the observed stochastic patterns 
of phase changes during pseudoelastic twinning. 
Therefore, tapping into the microscopic dynamics can offer
a more effective strategy to perform inference-based analyses of distributions, 
including those acquired experimentally.
In this article, we model and evaluate 
the \emph{critical} force responses associated with a prototypal martensitic transition
responsible for twinning.
We also predict the statistical first moment (mean) of the 
critical force distribution as a function of the imposed strain rate (or force rate),
which, to our knowledge, has yet to be experimentally measured.
As an outcome, we demonstrate how this response-stimulus correlation
reveals fundamental properties
of the phase transition. 

\begin{figure*}
\centerline{\includegraphics[width=\textwidth,trim= 0mm 0mm 0mm 0mm,clip]{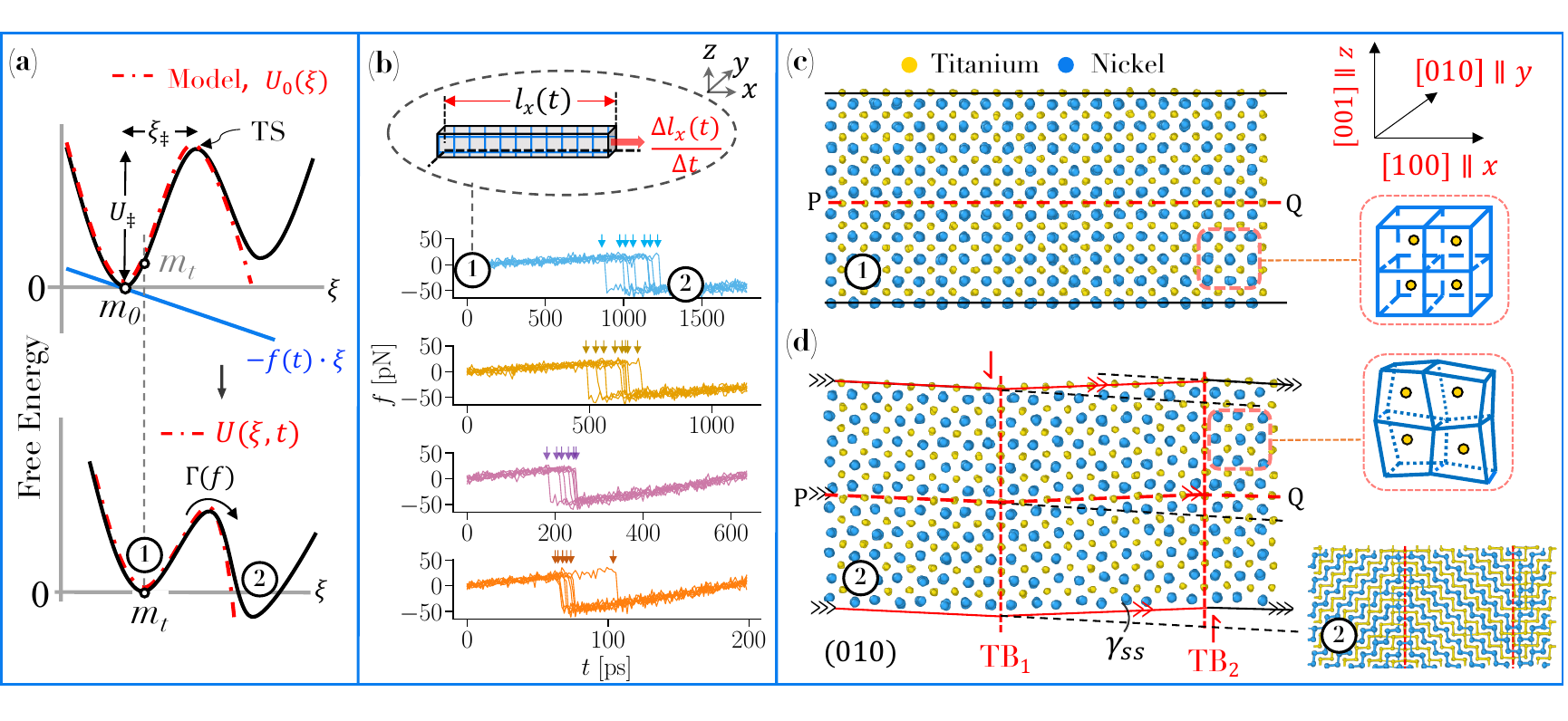} }
\caption{
{\bf Deformation twinning of titanium nickel single crystal. \label{fig1} }
(a)
(Top) Schematic of an equilibrium free energy profile (solid line) and a parameterised model, $U_0(\xi)$
(dot-dashed), used for reconstruction. The stationary state is denoted by $m_0 \equiv \xi_0$=0.
(Bottom) At a fixed temperature, an imposed force, $f(t)$, causes the profile to incline by $-f(t) . \xi$.
(b) Schematic of molecular dynamics simulation box under
constant uniaxial tensile deformation rate, ($\Delta l_x(t) / \Delta t$).
Force-time traces, top to bottom,
at ($\Delta l_x/ \Delta t$) [\AA/ps]:
(i)   3.679 $\times$ 10$^{-4}$,
(ii)  7.839 $\times$ 10$^{-4}$,
(iii) 2.352 $\times$ 10$^{-3}$ and
(iv)  7.525 $\times$ 10$^{-3}$. Short vertical arrows indicate phase transition events leading to lattice twinning.
(c) Atomic configuration of
\textcircled{\scriptsize{1}}---austenite phase, its projection on (010) and 
the coordinate system are shown. Viewing direction is along $y$--[010].
(d) Atomic configuration of twinned lattice \textcircled{\scriptsize{2}}.
Miller indices refer to the orthogonal axes system of the parent lattice.
Twin boundaries are denoted as TB$_1$ and TB$_2$, shear strain as $\gamma_{ss}$, and
a representative deformation is shown along PQ to highlight the change in 
atomic stacking relative to (c).
(Inset) Complexly patterned twinned martensite configuration, visualised as a display of bonds with lengths restricted to 3 \AA \, or less.
}
\end{figure*}

In the microscopic description,
the phase transition underlying twinning is represented on
a one-dimensional free energy profile.
The austenite--\textcircled{\scriptsize{1}}
and martensite--\textcircled{\scriptsize{2}} states are denoted as a pair of local
minima separated by a barrier; \fig{fig1}(a), Top.
Initially, under a no-force condition,
the system resides in the austenite state, which is metastable
under thermal fluctuations at room temperature.
However, the rate of the austenite (cubic unit cells) to martensite 
(monoclinic unit cells) transformation via barrier-crossing events remains negligible, 
reflecting infrequent turnover due to the large 
barrier size relative to the thermal energy. 
When imposed with a linearly increasing time-dependent mechanical perturbation 
(e.g., constant displacement rate or its equivalent force rate), the frequency of 
austenitic states crossing the barrier increases exponentially. 
To model the dynamics, a scheme is required that
describes the spatiotemporal nature of the random transition process
through a relationship between the probability density of
the instantaneous state population
and the probability flux of states across the time-dependent free energy barrier.
This evolution is modelled here according to a time-dependent probabilistic 
equation after Einstein-Smoluschowski \cite{risken,Hanggi1990}, 
providing a framework to understand the response relations 
originating from the phase transition.

Gradually ramping the externally applied force, $f(t)$, 
linearly over time, $t$, 
causes the initial 
state 
of austenite to shift away progressively in time.
The free energy profile at equilibrium, $U_0(\xi)$, where $\xi$ denotes displacement,
undergoes a downward tilt,
contributing to a depression of the barrier, $U_\ddag$; refer to \fig{fig1}(a)-Bottom.
So, the instantaneous profile $ U(\xi, t)$ is expressed by subtracting
$\xi \cdot f(t) $ from $U_0(\xi)$.
Suppose a harmonic potential approximation of $U_0(\xi)$ is considered at $\xi_0$=0,
with a vertical barrier of size 
$U_\ddag$ at $\xi = \xi_\ddag$. Then
one finds, for a very small, imposed force, \mbox{$f \ll (U_\ddag / \xi_\ddag) $},
a reduction of the original barrier from
$U_\ddag$ to $U_\ddag(f) \approx U_\ddag - f \xi_\ddag $.
The force-induced lowering of the barrier enhances 
the austenite to martensite transformation rate: 
$\Gamma(f) \propto  \exp[-U_\ddag(f)/k_B T] \propto \exp(f \xi_\ddag / k_B T) $.
The exponential rate relation formally emerges
if the model of free energy profile is incorporated for solving 
the Einstein-Smoluschowski equation \cite{risken,Hanggi1990}.
For large barriers and quasistatic imposed force rates, its solution
provides the Kramers \cite{Kramers1940} rate of escape,
yielding models for force-dependent rate of
phase transition mediated twinning, $\Gamma(f | \dot{f})$,
and critical force distribution, $p(f_\ddag | \dot{f})$,
and its first moment, $\langle f_\ddag\rangle({\dot{f}})$
at different levels of imposed force rates, $\dot{f}$,
capturing rate-dependent properties of the phase transition.
These models, coupled with 
statistical fitting procedures, allow an analysis of
response measurements
to draw estimates of the free energy profile,
twin domain size, intrinsic kinetic constant, and activation barrier
 of the solid-state structural phase transition.

In the next section, we describe the methods---
acquiring twinning response patterns using molecular dynamics simulations,
processing
twinning-associated observables, and applying statistical procedures such as maximum
likelihood estimation for interpreting the critical force distributions.
In the section on Results and Discussion, we review the microscopic theory,
 outlining the models of distribution and its first moment
relevant to the current study. Finally, we evaluate and discuss
the outcomes of the statistical applications, providing insights into
the free energy profile of the martensitic transition.

\section{Methods}

\subsection{Molecular Dynamics Simulations}

Classical molecular dynamics \cite{dfrenkel96,KO201790,Srinivasan2018} calculations on titanium nickel (Ti:Ni=1:1)
single crystals were performed
using a modified embedded-atom method for the second-nearest neighbour
interatomic potential \cite{Ko2015,ctcms}.
Along the $x$, $y$, and $z$ axes, the simulation box had lengths of 60 \AA, 30 \AA,
and 30 \AA, respectively, aligning with the [100], [010], and [001]
directions of the periodic supercell, composed of 20 $\times$ 10 $\times$ 10
cubic unit cells.
Periodic boundary conditions were chosen for all axes.
Simulations were conducted and visualised, respectively using LAMMPS \cite{Plimpton1995} and Ovito \cite{ovito}.

Nonequilibrium simulations were conducted to generate response force-time
and force-displacement traces.
During nonequilibrium simulations, the atomic system was evolved with a timestep of 1 fs within the NPT ensemble. The temperature was maintained at T=300 K, and pressures were constrained along the $y$ and $z$ axes at P=1.013 bar, employing the N\'{o}se-Hoover approach
\cite{Shinoda2004,Tuckerman2006}.
The damping times specified for temperature and pressure relaxation were 0.7 ps and 1 ps, respectively.
Thus, the system underwent relaxation within a few picoseconds,
many orders of magnitude faster in kinetics than the structural transformation itself.
The propagation of every tensile simulation had ensued
from a random atomic
configuration (microstate), which was sampled from the equilibrium (i.e., under no imposed force)
isothermal-isobaric distribution of the system,
prepared through 0.5-1 ns, to allow for its relaxation.

A homogeneous strain was applied to the simulation box, with the
box dimension along the $x$ axis, increasing
in equal increments every timestep.
The simulation trajectories and response traces were acquired at the following rates of uniaxial extension, ($\Delta l_x/\Delta t $),
[\AA/ps]:
(i) 9.03 $\times$ 10$^{-5}$,
(ii) 1.79 $\times$ 10$^{-4}$,
(iii) 3.679 $\times$ 10$^{-4}$,
(iv) 7.839 $\times$ 10$^{-4}$,
(v) 2.352 $\times$ 10$^{-3}$,
(vi) 7.525 $\times$ 10$^{-3}$ and
(vii) 2.35 $\times$ 10$^{-2}$.
The number of independent nonequilibrium simulations performed
for tensile extensional rates (i)-(ii) was 30, and (iii)-(vii) was 400.
The maximal strain was constrained within 2.5 \% for elastic twin generation.
At every timestep, the system's net force that was generated by the imposed deformation
 was calculated using  $f(t) = (A_{xx}/n) \sigma_{xx}(t)  $.
Here, $\sigma_{xx}$ represents the \emph{normal} stress component of the
internal stress tensor, defined through an expression of the system's virial stress
\cite{thompson2009general}, and ($n/A_{xx}$) is the number density of atoms
in the plane with a normal vector parallel to $x$ axis.
The timescales of the applied strain rates were much smaller than the internal relaxation
rate of the atomic system.
With this criterion, on average, the externally imposed force, $f_{ext}$, remained in balance
with the instantaneous force developed in the system, $\langle f(t) \rangle$,
that is, $f_{ext} = \langle f(t) \rangle $.

\subsection{Critical forces and normalised histograms }

The steps for retrieving critical forces and histograms 
from the molecular dynamics simulations are described.
The force-time traces serve as the primary data and
are processed to give the probability density functions (PDF)
of critical forces for a given deformation rate.
Representative samples of $f$-$t$ traces are displayed in \fig{fig1}(b).
For the extraction of PDFs,
the transition time, $t_\ddag$, at which peak force occurs (before the transition)
is identified, followed by its extraction from every $f$-$t$ trace, for all strain rates.
Then, the transition time is converted to force value by multiplying with
force-rate
[$\dot{f} = \partial \langle f(t) \rangle / \partial t$],
computed from the gradient of the averaged $f$-$t$ trace, denoted
as $\langle f(t) \rangle $,
in a small time interval near $t=0$.
Corresponding to uniaxial tensile deformation rates
($\Delta l_x / \Delta t$) imposed,
the following are the estimates of force rates $\dot{f}$, [pN/ps] observed:
(i) 3.246 $\times$ 10$^{-3}$,
(ii) 7.137 $\times$ 10$^{-3}$,
(iii) 1.482 $\times$ 10$^{-2}$,
(iv) 3.097 $\times$ 10$^{-2}$,
(v) 9.452 $\times$ 10$^{-2}$,
(vi) 2.927 $\times$ 10$^{-1}$, and
(vii) 9.242 $\times$ 10$^{-1}$.
In our simulations, we note that the phase transition event immediately leads to twinning.
Finally, the set of critical transition (or phase transition) forces $\{ f_{\ddag} | \dot{f} \}$
for a given $\dot{f}$ is translated into a normalised
histogram, $\widehat{p}(f | \dot{f})$, as shown in \fig{fig2}, and
its first moment (mean), $\langle f \rangle (\dot{f}) $, computed
in dependence of the measured force rates; \fig{fig3}.
The correlation $\langle f \rangle$ vs.\ $\dot{f} $ is referred to as the force spectrum.
The symbol $\wedge$ over a variable is used as a convention to 
distinguish that variable from its
estimated value.

\subsection{Estimation of parameterised landscape}

Estimates of the landscape parameters---the free energy barrier,
barrier location and intrinsic frequency of phase transition
were obtained independently from statistical fits of
the respective models of the PDF of critical pase transition force
 and the PDF's first moment versus imposed forcing rates.
In the first case, the PDF of the critical force, Eq.\ (6), was
fit to the response observable $\widehat{p}(f ; \dot{f})$ using the method of maximum likelihood estimation (MLE) \cite{hastie2009elements},
which is a statistical approach of parameter inference for models of distribution functions.
It allows the determination of the model landscape
that is most likely to have generated a random sample
of phase transition forces $\{ f_{ij} \} \equiv  f_{11}, f_{21}, \cdots, f_{ij}, \cdots $, which were
retrieved from the force response data at different levels
of strain rates or equivalent force rates, $\dot{f}$,
as applicable in the present case.
See section III.B for the form of the free energy profile.
The critical force is denoted as $f_{ij}$
for the $i$-th sample trace, measured at a force rate $\dot{f_j}$.
Here, $i = 1, \ldots , S$, where $S$=400
is the sample size
of the traces recorded for the
$j$-th force rate and $j=1, \ldots , N_r$, where $N_r$=4 is the
number of distinct strain rates used for performing MLE.
Denote the PDF of transition force vector $\{ f_{ij} \}$
as $p(\{ f_{ij} \} | \vec{w})$ for a
given parameter vector $\vec{w} \equiv \{U_\ddag, x_\ddag , \Gamma_0 \}$,
which defines the free energy profile.
Noting that each observation
$f_{ij}$ is statistically independent,
the PDF of $\{ f_{ij} \}$
can be expressed as a product of the PDF of individual observations:
$p( \{ f_{ij} \} | \vec{w}) = p(f_{11}| \vec{w}) p(f_{12}| \vec{w}) \ldots p(f_{SN_r}|\vec{w}) $.
According to MLE, a likelihood function $L_{mle}$ of the set of parameters $\vec{w}$, given a set $\{f_{ij} \}$,
is defined as:
\begin{equation}
L_{mle}(\vec{w} | \{f_{ij} \} ) = \prod_{j=1}^{N_r}  \prod_{i=1}^{S} p( f_{ij}  ; \vec{w}  )  . \label{lhood}
\end{equation}
An individual probability, $p(f_{ij}; \vec{w})$, is given by the model \eqn{fdistr},
while the products are computed over
the set of phase transition events observed, totalling
$S_j \times N_r $  = 1600 events.
Thus, the likelihood function evaluates how well a given landscape describes
a transition force distribution.
The log-likelihood, $ ( \ln L_{mle} )$, function was maximised by varying $\vec{w}$,
 using the Nelder-Mead nonlinear optimising
algorithm \cite{scipy,nelder1965simplex},
within a tolerance value on $( \ln L_{mle} )$, set
for achieving convergence of the algorithm, providing the most likely landscape parameters.
In the second case,
the model of first moment, Eq.\ (8),
was fit to the expectation value of the distributions, $ \langle \widehat{f} \rangle(\dot{f}) $, using least square \emph{regression}.
Minimisation of the sum of squared residuals (that is, differences)
between predicted and actual values
was achieved using the conjugate gradient algorithm, giving
optimal landscape parameters and standard errors (s.e); Table \ref{table1}.

\section{Results and Discussions}

\begin{figure*}[ht]
\centerline{\includegraphics[width=0.7\linewidth,trim= 0mm 0mm 0mm 0mm,clip]{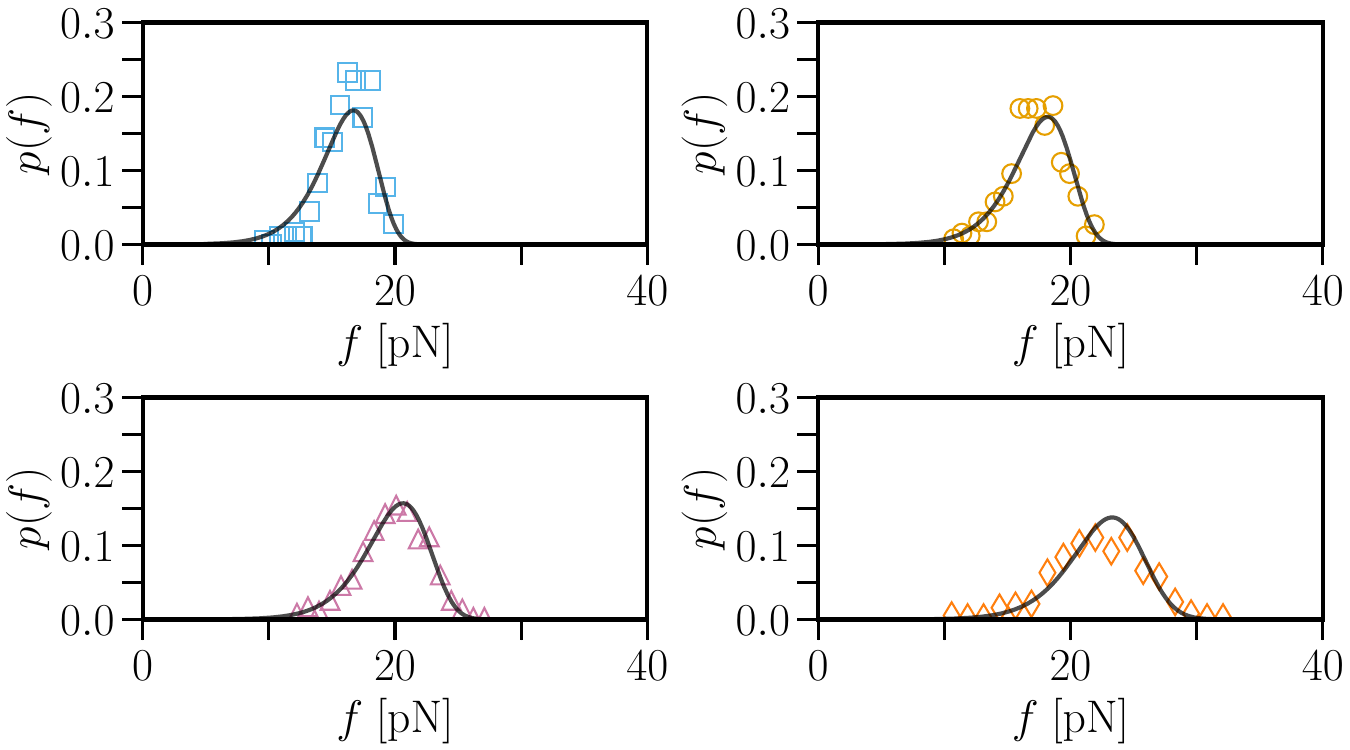} }
\caption{
{\bf Probability density distributions of critical force of martensitic phase transition. \label{fig2} }
Shown are the PDFs, $p(f)$, of phase transition forces $f_\ddag$ at deformation rates [\AA/ps]
(i)   3.679 $\times$ 10$^{-4}$ ($\square$),
(ii)  7.839 $\times$ 10$^{-4}$ ($\iscircle$),
(iii) 2.352 $\times$ 10$^{-3}$ ($\triangle$) and
(iv)  7.525 $\times$ 10$^{-3}$ ($\Diamond$).
Solid lines depict maximum likelihood estimated fits of \eqn{fdistr} using a single set of optimum parameters
given in Table \ref{table1}.
}
\end{figure*}

\subsection{
Critical force distribution is a defining characteristic of a transition mechanism}

Consistent with the behaviour expectation for an elastic solid,
linear stress-strain and force-time traces were initially observed
when subjecting single crystalline titanium nickel in the austenite (cubic) phase
to a constant uniaxial tensile extension rate.
This was followed by a sharp drop in stress or force
on encountering a phase transition,
leading to a change in symmetry of the lattice to a twinned martensite
phase, 
which served to relieve the stress in the system.
Representative atomic configurations, before and after the occurrence of
phase transition mediated twinning, are depicted in \fig{fig1}(c) and (d), respectively.
The traces of the twin boundaries TB$_1$ and TB$_2$, which coincide with the Ti atomic plane, 
are shown as vertical dashed lines parallel to (100). The unit cell variants on either side of a twin boundary
are related by reflection symmetry.
Further, representative faults in the stacking of (100) planes can be viewed along the
line PQ on plane 
(010) in the transformed configuration.
The martensite structure and twin planes
corroborate with 
the detailed crystallographic analysis of twinning in titanium nickel conducted by Li et al. \cite{Li2020} 
using an identical interatomic potential \cite{Ko2015} as this work.
We focus on resolving
the inherent stochasticity of the martensitic transformation evident
from the variations
in the critical phase transition forces observed over independent realisations
of the system's trajectory under identical conditions.
The characteristics of a transition mechanism are captured by
the critical force distribution, represented by
the conditional probability density, $\widehat{p}(f | \dot{f})$;
refer to Fig.\ 2.
To quantify the strain rate effects,
we compute the statistical first moment $\langle f_\ddag \rangle $ of the critical force
distribution acquired at the different force rates $\dot{f}$.
The $\langle f_\ddag \rangle $ vs.\ $\dot{f}$  plot displays a nonlinear relation
 with considerable variation of strength; refer to \fig{fig3}.
Interpreting the physical basis of twinning, as outlined in
$p(f | \dot{f})$ and $\langle f_\ddag \rangle(\dot{f})$,
necessitates models of transition observables.
Employing a microscopic theoretical approach, we illustrate how the response relations
express fundamental characteristics of the solid-state transformation.

\subsection{Quantitative description of phase transition linked force distribution}

The phase transition can be modelled using a mechanistic formulation
wherein the dynamics proceeds 
on a one-dimensional free energy landscape, $U_0(\xi)$; \fig{fig1}(a).
The state of the crystal is denoted by $\xi$,
an order parameter to track the change of state.
In this perspective, atomic configurations pass
from state \textcircled{\scriptsize{1}} through a transition state (TS)
 to state \textcircled{\scriptsize{2}}.
The TS acts as a barrier to the transition flux of the states
since it has a higher free energy than both the parent and product configurations; see \fig{fig1} (Top).
The imposed time-dependent deformation, increasing monotonically in time,
translates into increasing force levels and transition probabilities.
The imposed force modifies the equilibrium profile,
$U_0(\xi)$, such that $U(\xi, t) \approx U_0(\xi) - \langle f(t) \rangle \, \xi$.
Here, $\langle f(t)  \rangle$  denotes a statistical mean of the instantaneous force developed
in the atomic system. It is computed as an average over the set
of experiments or simulations performed at a constant rate of uniaxial
tensile deformation:
\begin{equation}
\langle f(t) \rangle \approx \dot{f} t \equiv f(t). \label{linforcetime}
\end{equation}
The external perturbation
causes the free energy landscape to slant.
In addition, the curvature of the landscape and
 locations of free energy minima and the TS at $\xi = \xi_\ddag$ are concurrently modified.
These changes enable a faster turnover rate of the austenite to martensite,
making the latter a preferred physical state.

For analytical modelling, consider
a free energy function \cite{Garg1995}:
$U_0(\xi) = U_\ddag/2 + (3U_\ddag/2\xi_\ddag) ( \xi -  \xi_\ddag/2) - (2 U_\ddag/(\xi_\ddag)^3) (\xi - \xi_\ddag/2)^3$, where
$U_\ddag$ is the free energy barrier (under no force condition), given as
a difference in the free energy of TS and
the free energy minimum of \textcircled{\scriptsize{1}}, and
$\xi_\ddag$ is the distance of the TS from the position ($m_0$) of
free energy minimum of state \textcircled{\scriptsize{1}}.
%
According to Einstein-Smoluschowski equation, the state probability density function, $P(\xi, t)$, is governed by
$ \partial_t P(\xi, t) = -\partial_\xi [ - (k_B T/ \eta) \, \partial_\xi P(\xi, t) + (-\partial_\xi U(\xi, t) / \eta ) \, P(\xi, t) ]$
\cite{risken,Hanggi1990}, for the case of overdamped thermal activation.
The equation assumes that the inertial term ($\ddot{\xi}$) is negligible
based on strong drag forces captured via a frictional coupling, $\eta$,
contributing to the dissipation of configurational fluctuations.
An overview of the analytical
solutions \cite{risken,Hanggi1990,Kramers1940,Garg1995,Dudko2003}
is presented.
The differential equation is simplified in terms of survival probability:
$\Psi(t) = \int_{-\infty}^{\xi_\ddag} P(\xi, t) d\xi$,
defined as a cumulative distribution of the probability density
of system state observed at time $t$ and domain, $-\infty < \xi < \xi_\ddag $.
$\Psi(t)$ gives the probability of austenite state to persist
till time $t$.
Assuming steady-state flux across the barrier,
and the boundary at barrier position is absorbing,
and further replacing variable $t$ with $f(t)$ via \eqn{linforcetime},
the evolution equation can be recast into
\begin{equation}
-\partial_f \Psi(f) = \left [ \Gamma(f) \Psi(f) \right ] / \dot{f}  . \label{diffeqn}
\end{equation}
$\Gamma(f)$ is the force-dependent rate of austenite to martensite transformation:
\begin{equation}
\Gamma(f)  \approx \gamma(f) \; { \exp \left (\Delta U_\ddag(f)/k_B T \right ) },   \\ \label{gammaf}
\end{equation}
the force-dependent terms of
$ \gamma(f) = \Gamma_0 \left ( 1 - (f/f_c) \right )^{1/2} $ is the
attempt frequency of atomic configurations to change state and
$\Delta U_\ddag (f) = { U_\ddag  \left ( 1 - \left \{ 1 - (f/f_c)  \right \}^{3/2} \right ) }$
is the change in barrier with force $f$.
The maximal force to turnover austenitic state is $f_c \equiv (3U_\ddag/2\xi_\ddag )$.
The prefactor in the attempt frequency, at $f$=0, is given as
$\Gamma_0 = \{ 1/(2 \pi \eta ) \} \cdot (6U_\ddag/{\xi_\ddag}^2) \;  e^{-U_\ddag/kT}$.
The survival probability, $\psi(f)$, derived from
\eqn{diffeqn} is
\begin{equation}
\psi(f) =   \exp {  \left \{ \mu_0 - \mu(f)  \left ( 1 - (f/f_c) \right )^{-1/2} \right \}} , \label{surv}
\end{equation}
where $\mu(f) \equiv (\Gamma(f) \, k_B T) / ( \dot{f} \, \xi_\ddag )  $.
As the imposed force increases,
the survival probability of austenite decreases rapidly,
from $\psi$($f$=0)=1, approaching zero
as $f$ approaches $f_c$,
since the austenite microstates
progressively switch to martensite.
The model PDF is obtained
by substituting \eqn{surv} in
$p(f | \dot{f}) = - \partial_f \psi $:
\begin{equation}
p(f \, | \, \dt{f}) = \left \{ \frac{\Gamma(f) }{ \dt{f}} \right \}  \exp { \left \{ \mu_0 - \mu(f)  \left ( 1 - (f/f_c) \right )^{-1/2} \right \}}. \label{fdistr}
\end{equation}
Here, $\mu_0 \equiv \mu(f=0)$ is the rate $( (k_B T) \cdot \Gamma_0 )$
of thermal energy transfer across the
barrier at equilibrium, expressed as a fraction of the
absorption rate of lattice strain energy, $( \xi_\ddag  \dot{f} )$,
due to the imposed forcing.
The analytical model, \eqn{fdistr}, along with maximum likelihood estimation,
enables the deduction of free energy profile, interaction range, and
intrinsic kinetics governing the phase transition linked to the twinning mode of deformation.

\subsection{Nonlinear relationship between critical force and imposed loading rate}
\begin{figure}
\centerline{\includegraphics[width=\linewidth,trim= 0mm 0mm 0mm 0mm,clip]{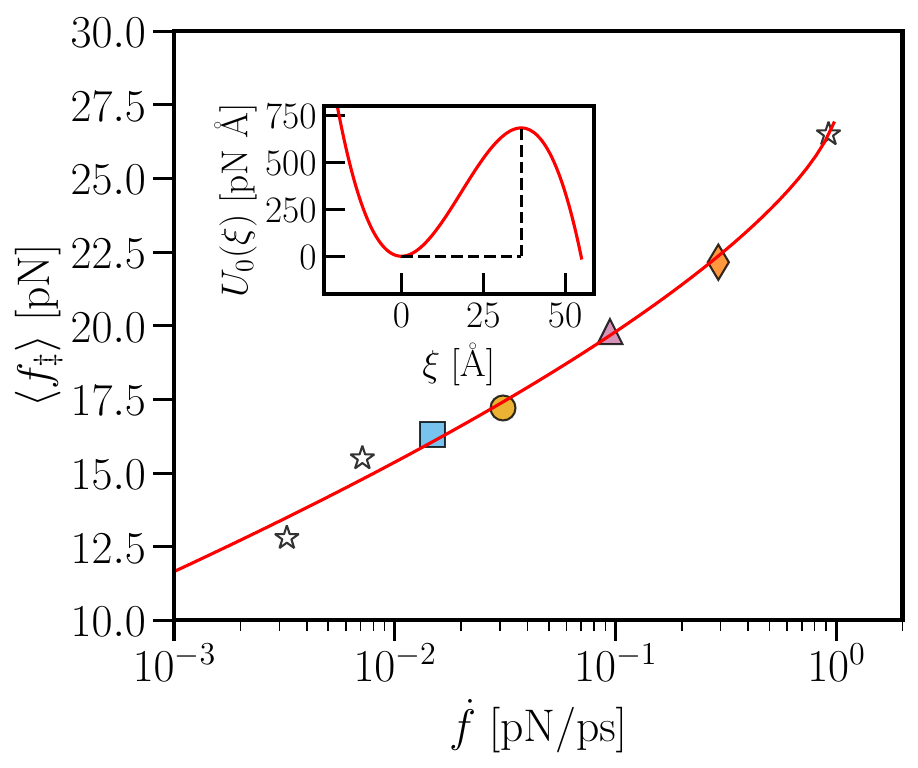} }
\caption{{\bf Mean critical force of martensitic transition vs force rate.\label{fig3} }
First moment of $p(f | \dot{f})$, at different rates of tensile force rates $\dot{f}$ [pN/ps].
(Solid line) Least square regression model, \eqn{avgf}.
(Inset) Predicted free energy profile using best-fit parameters of \eqn{avgf} given in Table \ref{table1}. }
\end{figure}
\fig{fig3} (symbols) shows the relation between mean transition force to twin a lattice
and the imposed force rate obtained from simulations.
It shows a higher strength of the lattice (austenite) in the fast loading rate regime
and lower strength at slower loading rates.
In quantitative terms, a doubling of strength correlates with an approximately 
300-fold increase in the force rate.
It can be explained from
the shift in energy minimum initially positioned on $m_0$ = 0 to the right, $m_t$, a configuration
with a higher deformation associated landscape force, as the landscape leans downwards; \mbox{\fig{fig1}(a) and (b)}.
Consequently, under external perturbation, most configurations escape
the free energy well beyond the value $\xi = m_t$ instead of the equilibrium value $\xi = m_0$.
Since the landscape force $f(\xi = m_t)$ is larger than $f(\xi = m_0)$, this translates, on average,
a higher force requirement for the escape of the atomic configurations
still extant in the austenitic state.

The relation between the expectation value of the phase transition force $\langle f_\ddag \rangle $ vs.\ $\dot{f}$
can be derived as follows \cite{Friddle2008a}.
Integration
of
Eq.\ \eqref{diffeqn},
$\dot{f} \int_1^\psi d\psi/\psi =-\int_0^{f} \Gamma(f) df$, after substituting \eqn{gammaf}
gives:
\begin{equation}\label{f1s}
f(\omega)  = f_c \left [ 1 - \left \{ 1 - \omega \right \}^{2/3} \right ] ,
\end{equation}
where $\omega(\psi) = (U_\ddag/(k_B T))^{-1} \ln \left(1 - \mu_0 \ln \psi \right)$.
Then, an expression for the expectation value (of the transition force)
can be evaluated by a series expansion of $f(\omega)$ in the vicinity of
$\omega = \langle \omega \rangle$ where $ \langle \omega \rangle = \int_0^1 \omega(\psi) d \psi$.
This step is followed by performing an ensemble average of the series expansion:
$\langle f \rangle = f(\langle \omega \rangle) + (1/2) \langle  (\omega - \langle \omega \rangle)^2 \rangle  f''(\langle \omega \rangle) + \cdots \approx f(\langle \omega \rangle)  $.
The higher-order terms
$f''(\langle w \rangle)$ and beyond are small
and neglected. Finally, the required expression is obtained
\begin{equation}
\langle f \rangle   \cong f_c \left [ 1 - \left \{ 1 - e^{\mu_0} E_1(\mu_0) / (U_\ddag/(k_B T)) \right  \}^{2/3}  \right] \equiv \langle f_\ddag \rangle  \label{avgf}\\
\end{equation}
where  $E_1(\mu_0) = \int_{\mu_0}^\infty \frac{e^{-z}}{z} dz$ is an exponential integral \cite{AbramSteg}.
The physical basis of this transition law is thus a consequence
of the time-dependent remodelling of the positions of austenite's
local minimum and transition state and the height of
 the barrier during the time course of deformation.

\subsection{Response relations reflect fundamental properties of
solid-state transformation-mediated twinning}

The strain rate effects of
critical phase transition force distribution and its spectra
collected from molecular dynamics simulations are characterised
using the statistical mechanics-based models described in the previous subsections.
The PDF \eqn{fdistr} was optimised via the maximum likelihood estimation method
 to fit the simulated distributions.
The model with optimum parameters
is plotted for all strain rates in \fig{fig2}, solid lines.
Whereas the line of regression for \eqn{avgf},
which captures the simulated pattern,
$\langle f \rangle(\dot{f})$,
is plotted as a solid line in \fig{fig3}.
The values of the optimal model parameters and
standard errors (s.e) for both cases are given in Table \ref{table1}.
Both models
effectively capture the variability observed in each
simulated response type and across loading rates.
The microscopic significance of the parameters is as follows.
The estimate of $U_\ddag$ suggests the stacking-fault
free energy activation requirement for twin formation.
Specifically, an average value of
($U_\ddag / a^2)$ involving 20 atomic layers of (100) planes yields
$\approx$ 34 mJ/m$^2$ ($a \approx$ 3 \AA \, is lattice parameter), which
lies in the range of estimates reported
from density functional theory calculations
\cite{Chowdhury2017}.
In addition,
$\xi_\ddag$ and $\Gamma_0$, respectively,
are estimates of the domain size of nanotwin produced
and the intrinsic frequency of converting
atomic configurations (unit cells) from the austenite to martensite phases.
While the models demonstrate high explanatory performance, our focus has been on the
material response under fast loading rates or strain rates greater than 10$^{-4}$ \AA/ps.
Further scrutiny
at lower force rates is required
to establish their applicability for analyses of experimental responses.
Nonetheless, the study addresses the nature of
response-stimulus laws, utilising statistical mechanics to
illustrate how such laws reflect the thermodynamic and kinetic features
of the underlying phase change in the solid state.
\begin{table}
\centering
\caption{ Estimates of free energy landscape parameters for martensitic transition 
in titanium nickel from statistical fits of Eqs.\ (6), and (8) to the respective deformation response data $\widehat{p}(f | \dot{f})$ and $\langle f \rangle ( \dot{f} )$ .
Standard errors (s.e.) of parameters are specified in parenthesis. } \label{table1}
\begin{center}
\begin{tabularx}{\linewidth}{ | X | X | X |  }
 \hline
 { Parameters}              &  $p(f | \dot{f})$  &  $\langle f \rangle ( \dot{f} )$ \\
 \hline
 $ U_{\ddag} $, pN \AA         & 638.14          		 & 659    \\
 $ \pm ( \mathrm{s.e.} ) $     & (9.62)          		 & (38.6)    \\
 \hline
 $ \xi_{\ddag} $, \AA          & 29.2 			& 36.1  \\
 $ \pm ( \mathrm{s.e.} ) $     & (0.45)			& (3.6)  \\
 \hline
 $ \Gamma_{0} $, 1/ps          &   3.96 $\times$ 10$^{-7}$    &  6.12 $\times$ 10$^{-8}$  \\
 $ \pm ( \mathrm{s.e.} ) $     &   (6.75 $\times$ 10$^{-8}$)  &  (6 $\times$ 10$^{-8}$) \\
 \hline
\end{tabularx}
\end{center}
\end{table}

\vspace{0.5cm}
\section{Conclusion}

Dynamic nanoscale force profiles 
allow probing 
variations in
atomic scale interactions within crystal structures,
offering fundamental insights into phase 
transformation kinetics under perturbation. 
In the context of the phase transition-mediated deformation 
twinning process in single crystalline titanium nickel, 
microscopic simulations indicate how systematic analyses of \emph{stochasticity} in
force responses at constant loading rates can unravel some previously 
unexplored response-stimulus laws of mechanical origin. 
The critical-force distributions, $p(f_\ddag | \dot{f})$, 
and the characteristic correlation 
of the first moment, $\langle f_\ddag \rangle$,
vs.\ the imposed tensile force rate, $\dot{f}$, are predicted. 
A nonequilibrium statistical mechanics framework based on overcoming a free energy barrier via random structural transitions explains the observed patterns. 
The models in this study have strong explanatory power, but further
validation is needed, especially for the low strain rates
that were not examined.
Even so, 
the framework holds potential beyond simulated data.
The observables provide targets for experimental nanomechanical measurements, 
as, to our knowledge, they have not yet been experimentally determined.
This can pave the way for new studies 
on solid-state structural transitions in various other metals and
for reconstructing their free energy profiles employing nanomechanical techniques. 
%
%

\noindent \emph{Acknowledgement:} The authors gratefully acknowledge
the use of high performance computing facility and IT resources at BMU.


\begin{thebibliography}{59}
\expandafter\ifx\csname url\endcsname\relax
  \def\url#1{\texttt{#1}}\fi
\expandafter\ifx\csname urlprefix\endcsname\relax\def\urlprefix{URL }\fi
\expandafter\ifx\csname href\endcsname\relax
  \def\href#1#2{#2} \def\path#1{#1}\fi

\bibitem{Dehm2018}
G.~Dehm, B.~N. Jaya, R.~Raghavan, C.~Kirchlechner, {Overview on micro and
  nanomechanical testing: New insights in interface plasticity and fracture at
  small length scales}, Acta Mater. 142 (2018) 248--282.
\newblock \href {https://doi.org/10.1016/j.actamat.2017.06.019}
  {\path{doi:10.1016/j.actamat.2017.06.019}}.

\bibitem{Kiener2024}
D.~Kiener, A.~Misra, {Nanomechanical characterization}, MRS Bull. 49~(3) (2024)
  214--223.
\newblock \href {https://doi.org/10.1557/s43577-023-00643-z}
  {\path{doi:10.1557/s43577-023-00643-z}}.

\bibitem{Dimiduk2005}
D.~M. Dimiduk, M.~D. Uchic, T.~A. Parthasarathy, {Size-affected single-slip
  behavior of pure nickel microcrystals}, Acta Mater. 53~(15) (2005)
  4065--4077.
\newblock \href {https://doi.org/10.1016/j.actamat.2005.05.023}
  {\path{doi:10.1016/j.actamat.2005.05.023}}.

\bibitem{FRICK2008}
C.~Frick, B.~Clark, S.~Orso, A.~Schneider, E.~Arzt, Size effect on strength and
  strain hardening of small-scale [111] nickel compression pillars, Materials
  Science and Engineering: A 489~(1) (2008) 319--329.
\newblock \href {https://doi.org/https://doi.org/10.1016/j.msea.2007.12.038}
  {\path{doi:https://doi.org/10.1016/j.msea.2007.12.038}}.

\bibitem{Binning}
G.~Binnig, H.~Rohrer, C.~Gerber, E.~Weibel, Surface studies by scanning
  tunneling microscopy, Phys. Rev. Lett. 49 (1982) 57--61.
\newblock \href {https://doi.org/10.1103/PhysRevLett.49.57}
  {\path{doi:10.1103/PhysRevLett.49.57}}.

\bibitem{Zhong2024}
L.~Zhong, Y.~Zhang, X.~Wang, T.~Zhu, S.~X. Mao, {Atomic-scale observation of
  nucleation- and growth-controlled deformation twinning in body-centered cubic
  nanocrystals}, Nat. Commun. 15~(1) (2024) 1--9.
\newblock \href {https://doi.org/10.1038/s41467-024-44837-8}
  {\path{doi:10.1038/s41467-024-44837-8}}.

\bibitem{Bhowmick2019}
S.~Bhowmick, H.~Espinosa, K.~Jungjohann, T.~Pardoen, O.~Pierron, {Advanced
  microelectromechanical systems-based nanomechanical testing: Beyond stress
  and strain measurements}, MRS Bull. 44~(6) (2019) 487--493.
\newblock \href {https://doi.org/10.1557/mrs.2019.123}
  {\path{doi:10.1557/mrs.2019.123}}.

\bibitem{Zhu2005}
Y.~Zhu, H.~D. Espinosa, An electromechanical material testing system for <i>in
  situ</i> electron microscopy and applications, Proceedings of the National
  Academy of Sciences 102~(41) (2005) 14503--14508.
\newblock \href {https://doi.org/10.1073/pnas.0506544102}
  {\path{doi:10.1073/pnas.0506544102}}.

\bibitem{Oliver1992}
W.~C. Oliver, G.~M. Pharr, {An improved technique for determining hardness and
  elastic modulus using load and displacement sensing indentation experiments},
  J. Mater. Res. 7~(6) (1992) 1564--1583.
\newblock \href {https://doi.org/10.1557/JMR.1992.1564}
  {\path{doi:10.1557/JMR.1992.1564}}.

\bibitem{Schuh2006}
C.~A. Schuh, {Nanoindentation studies of materials}, Mater. Today 9~(5) (2006)
  32--40.
\newblock \href {https://doi.org/10.1016/S1369-7021(06)71495-X}
  {\path{doi:10.1016/S1369-7021(06)71495-X}}.

\bibitem{Kaushik2022}
N.~C. Kaushik, A.~Maitra, J.~A. Vamsi, T.~S. Krishna, A.~T. Satya,
  {Understanding elastic/plastic nature of phases in Fe–13Cr–1C hardfaced
  coating through accelerated property mapping technique}, Mater. Lett.
  320~(April) (2022) 132335.

\bibitem{uchic2004}
M.~D. Uchic, D.~M. Dimiduk, J.~N. Florando, W.~D. Nix, {Sample Dimensions
  Influence Strength and Crystal Plasticity}, Science 305~(5686) (2004)
  986--989.
\newblock \href {https://doi.org/10.1126/science.1098993}
  {\path{doi:10.1126/science.1098993}}.

\bibitem{Schuh2004}
C.~A. Schuh, A.~C. Lund, {Application of nucleation theory to the rate
  dependence of incipient plasticity during nanoindentation}, J. Mater. Res.
  19~(7) (2004) 2152--2158.
\newblock \href {https://doi.org/10.1557/JMR.2004.0276}
  {\path{doi:10.1557/JMR.2004.0276}}.

\bibitem{Morris2011}
J.~R. Morris, H.~Bei, G.~M. Pharr, E.~P. George, Size effects and stochastic
  behavior of nanoindentation pop in, Phys. Rev. Lett. 106 (2011) 165502.
\newblock \href {https://doi.org/10.1103/PhysRevLett.106.165502}
  {\path{doi:10.1103/PhysRevLett.106.165502}}.

\bibitem{Friedman2012}
N.~Friedman, A.~T. Jennings, G.~Tsekenis, J.-Y. Kim, M.~Tao, J.~T. Uhl, J.~R.
  Greer, K.~A. Dahmen, Statistics of dislocation slip avalanches in nanosized
  single crystals show tuned critical behavior predicted by a simple mean field
  model, Phys. Rev. Lett. 109 (2012) 095507.
\newblock \href {https://doi.org/10.1103/PhysRevLett.109.095507}
  {\path{doi:10.1103/PhysRevLett.109.095507}}.

\bibitem{Pattamatta2014}
S.~Pattamatta, R.~S. Elliott, E.~B. Tadmor, {Mapping the stochastic response of
  nanostructures}, Proc. Natl. Acad. Sci. U. S. A. 111~(17) (2014).
\newblock \href {https://doi.org/10.1073/pnas.1402029111}
  {\path{doi:10.1073/pnas.1402029111}}.

\bibitem{Otsuka2005}
K.~Otsuka, X.~Ren, {Physical metallurgy of Ti-Ni-based shape memory alloys},
  Prog. Mater. Sci. 50~(5) (2005) 511--678.
\newblock \href {https://doi.org/10.1016/j.pmatsci.2004.10.001}
  {\path{doi:10.1016/j.pmatsci.2004.10.001}}.

\bibitem{Christian1995}
J.~W. Christian, S.~Mahajan, {Deformation twinning}, Prog. Mater. Sci. 39
  (1995) 1--157.

\bibitem{Beyerlein2014}
I.~J. Beyerlein, X.~Zhang, A.~Misra, {Growth twins and deformation twins in
  metals}, Annu. Rev. Mater. Res. 44~(May) (2014) 329--363.
\newblock \href {https://doi.org/10.1146/annurev-matsci-070813-113304}
  {\path{doi:10.1146/annurev-matsci-070813-113304}}.

\bibitem{Bhadeshia_geometry}
H.~Bhadeshia, Geometry of Crystals, Polycrystals, and Phase Transformations,
  CRC Press, 2018.

\bibitem{Uttam2020}
P.~Uttam, V.~Kumar, K.~H. Kim, A.~Deep, {Nanotwinning: Generation, properties,
  and application}, Mater. Des. 192 (2020) 108752.
\newblock \href {https://doi.org/10.1016/j.matdes.2020.108752}
  {\path{doi:10.1016/j.matdes.2020.108752}}.

\bibitem{Bhattacharya2004}
K.~Bhattacharya, S.~Conti, G.~Zanzotto, J.~Zimmer, {Crystal symmetry and the
  reversibility of martensitic transformations}, Nature 428~(6978) (2004)
  55--59.
\newblock \href {https://doi.org/10.1038/nature02378}
  {\path{doi:10.1038/nature02378}}.

\bibitem{Bhattacharya2005}
K.~Bhattacharya, R.~D. James, {The material is the machine}, Science 307~(5706)
  (2005) 53--54.
\newblock \href {https://doi.org/10.1126/science.1100892}
  {\path{doi:10.1126/science.1100892}}.

\bibitem{MohdJani2014}
J.~M. Jani, M.~Leary, A.~Subic, M.~A. Gibson, {A review of shape memory alloy
  research, applications and opportunities}, Mater. Des. 56 (2014) 1078--1113.
\newblock \href {https://doi.org/10.1016/j.matdes.2013.11.084}
  {\path{doi:10.1016/j.matdes.2013.11.084}}.

\bibitem{McCracken2020}
J.~M. McCracken, B.~R. Donovan, T.~J. White, {Materials as Machines}, Adv.
  Mater. 32~(20) (2020) 1--48.
\newblock \href {https://doi.org/10.1002/adma.201906564}
  {\path{doi:10.1002/adma.201906564}}.

\bibitem{Chowdhury2017}
P.~Chowdhury, H.~Sehitoglu, {Deformation physics of shape memory alloys –
  Fundamentals at atomistic frontier}, Prog. Mater. Sci. 88 (2017) 49--88.
\newblock \href {https://doi.org/10.1016/j.pmatsci.2017.03.003}
  {\path{doi:10.1016/j.pmatsci.2017.03.003}}.

\bibitem{Niitsu2020}
K.~Niitsu, H.~Date, R.~Kainuma, {Thermal activation of stress-induced
  martensitic transformation in Ni-rich Ti-Ni alloys}, Scr. Mater. 186 (2020)
  263--267.
\newblock \href {https://doi.org/10.1016/j.scriptamat.2020.05.010}
  {\path{doi:10.1016/j.scriptamat.2020.05.010}}.

\bibitem{Yan2016}
X.~Yan, P.~Sharma, {Time-Scaling in Atomistics and the Rate-Dependent
  Mechanical Behavior of Nanostructures}, Nano Lett. 16~(6) (2016) 3487--3492.
\newblock \href {https://doi.org/10.1021/acs.nanolett.6b00117}
  {\path{doi:10.1021/acs.nanolett.6b00117}}.

\bibitem{Kumar2020}
P.~Kumar, U.~V. Waghmare, First-principles phonon-based model and theory of
  martensitic phase transformation in niti shape memory alloy, Materialia 9
  (2020) 100602.
\newblock \href {https://doi.org/https://doi.org/10.1016/j.mtla.2020.100602}
  {\path{doi:https://doi.org/10.1016/j.mtla.2020.100602}}.

\bibitem{Ogata2005}
S.~Ogata, J.~Li, S.~Yip, {Energy landscape of deformation twinning in bcc and
  fcc metals}, Phys. Rev. B - Condens. Matter Mater. Phys. 71~(22) (2005)
  1--11.
\newblock \href {https://doi.org/10.1103/PhysRevB.71.224102}
  {\path{doi:10.1103/PhysRevB.71.224102}}.

\bibitem{Hatcherprb2009}
N.~Hatcher, O.~Y. Kontsevoi, A.~J. Freeman, Martensitic transformation path of
  niti, Phys. Rev. B 79 (2009) 020202.
\newblock \href {https://doi.org/10.1103/PhysRevB.79.020202}
  {\path{doi:10.1103/PhysRevB.79.020202}}.

\bibitem{GudaVishnu2010}
K.~{G Vishnu}, A.~Strachan, {Phase stability and transformations in NiTi from
  density functional theory calculations}, Acta Mater. (2010).
\newblock \href {https://doi.org/10.1016/j.actamat.2009.09.019}
  {\path{doi:10.1016/j.actamat.2009.09.019}}.

\bibitem{Zarkevich2014}
N.~A. Zarkevich, D.~D. Johnson, {Shape-memory transformations of NiTi:
  Minimum-energy pathways between austenite, martensites, and kinetically
  limited intermediate states}, Phys. Rev. Lett. 113~(26) (2014) 10797114.
\newblock \href {https://doi.org/10.1103/PhysRevLett.113.265701}
  {\path{doi:10.1103/PhysRevLett.113.265701}}.

\bibitem{Li2020}
B.~Li, Y.~Shen, Q.~An, {Structural origin of reversible martensitic
  transformation and reversible twinning in NiTi shape memory alloy}, Acta
  Mater. 199 (2020) 240--252.
\newblock \href {https://doi.org/10.1016/j.actamat.2020.08.039}
  {\path{doi:10.1016/j.actamat.2020.08.039}}.

\bibitem{Tang2018}
X.~Z. Tang, Q.~Zu, Y.~F. Guo, {The surface nucleation of tension twin via
  pure-shuffle mechanism: The energy landscape sampling and dynamic
  simulations}, J. Appl. Phys. 123~(20) (2018).
\newblock \href {https://doi.org/10.1063/1.5022880}
  {\path{doi:10.1063/1.5022880}}.

\bibitem{Muller2001}
I.~M{\"{u}}ller, S.~Seelecke, {Thermodynamic aspects of shape memory alloys},
  Math. Comput. Model. 34~(12-13) (2001) 1307--1355.
\newblock \href {https://doi.org/10.1016/S0895-7177(01)00134-0}
  {\path{doi:10.1016/S0895-7177(01)00134-0}}.

\bibitem{Falk1980}
F.~Falk, {Model free energy, mechanics, and thermodynamics of shape memory
  alloys}, Acta Metall. 28~(12) (1980) 1773--1780.
\newblock \href {https://doi.org/10.1016/0001-6160(80)90030-9}
  {\path{doi:10.1016/0001-6160(80)90030-9}}.

\bibitem{Falk1983}
F.~Falk, {Ginzburg-Landau theory of static domain walls in shape-memory
  alloys}, Z. Phys. B 51 (1983) 177--185.

\bibitem{Maitra2022}
A.~Maitra, B.~Singh, {Interpreting force response patterns of a mechanically
  driven crystallographic phase transition}, Phys. Rev. Mater. 6~(4) (2022)
  043404.
\newblock \href {https://doi.org/10.1103/PhysRevMaterials.6.043404}
  {\path{doi:10.1103/PhysRevMaterials.6.043404}}.

\bibitem{risken}
H.~Risken, The Fokker‐Planck‐Equation. Methods of Solution and
  Applications, Springer-Verlag, Berlin, 1989.

\bibitem{Hanggi1990}
P.~Hanggi, P.~Talkner, M.~Borkovec, {Reaction-rate theory: fifty years after
  Kramers}, Rev. Mod. Phys. 62~(2) (1990) 251--341.

\bibitem{Kramers1940}
H.~Kramers, {Brownian motion in a field of force and the diffusion model of
  chemical reactions}, Physica 7~(4) (1940) 284--304.
\newblock \href {https://doi.org/10.1016/S0031-8914(40)90098-2}
  {\path{doi:10.1016/S0031-8914(40)90098-2}}.

\bibitem{dfrenkel96}
D.~Frenkel, B.~Smit, Understanding Molecular Simulation: From Algorithms to
  Applications, 2nd Edition, Vol.~1 of Computational Science Series, Academic
  Press, San Diego, 2002.

\bibitem{KO201790}
W.-S. Ko, S.~B. Maisel, B.~Grabowski, J.~B. Jeon, J.~Neugebauer, Atomic scale
  processes of phase transformations in nanocrystalline niti shape-memory
  alloys, Acta Materialia 123 (2017) 90--101.
\newblock \href {https://doi.org/https://doi.org/10.1016/j.actamat.2016.10.019}
  {\path{doi:https://doi.org/10.1016/j.actamat.2016.10.019}}.

\bibitem{Srinivasan2018}
P.~Srinivasan, L.~Nicola, A.~Simone, {Atomistic modeling of the
  orientation-dependent pseudoelasticity in NiTi: Tension, compression, and
  bending}, Comput. Mater. Sci. 154~(July) (2018) 25--36.
\newblock \href {https://doi.org/10.1016/j.commatsci.2018.07.028}
  {\path{doi:10.1016/j.commatsci.2018.07.028}}.

\bibitem{Ko2015}
W.~S. Ko, B.~Grabowski, J.~Neugebauer, {Development and application of a Ni-Ti
  interatomic potential with high predictive accuracy of the martensitic phase
  transition}, Phys. Rev. B. 92 (2015) 134107.
\newblock \href {https://doi.org/10.1103/PhysRevB.92.134107}
  {\path{doi:10.1103/PhysRevB.92.134107}}.

\bibitem{ctcms}
{Hale, L and Trautt, Z and Becker, C}, Interatomic potentials repository,
  \url{https://www.ctcms.nist.gov/potentials/}, Last accessed on 01 Feb 2024
  (2018).

\bibitem{Plimpton1995}
S.~Plimpton, {Fast parallel algorithms for short-range molecular dynamics}, J.
  Comput. Phys. 117 (1995) 1--19.
\newblock \href {https://doi.org/10.1006/jcph.1995.1039}
  {\path{doi:10.1006/jcph.1995.1039}}.

\bibitem{ovito}
A.~Stukowski, {Visualization and analysis of atomistic simulation data with
  OVITO-the Open Visualization Tool}, {Modelling and Simulation in Materials
  Science and Engineering} {18}~({1}) ({JAN} {2010}).
\newblock \href {https://doi.org/{10.1088/0965-0393/18/1/015012}}
  {\path{doi:{10.1088/0965-0393/18/1/015012}}}.

\bibitem{Shinoda2004}
W.~Shinoda, M.~Shiga, M.~Mikami, {Rapid estimation of elastic constants by
  molecular dynamics simulation under constant stress}, Phys. Rev. B - Condens.
  Matter Mater. Phys. 69~(13) (2004) 16--18.
\newblock \href {https://doi.org/10.1103/PhysRevB.69.134103}
  {\path{doi:10.1103/PhysRevB.69.134103}}.

\bibitem{Tuckerman2006}
M.~E. Tuckerman, J.~Alejandre, R.~L{\'{o}}pez-Rend{\'{o}}n, A.~L. Jochim, G.~J.
  Martyna, {A Liouville-operator derived measure-preserving integrator for
  molecular dynamics simulations in the isothermal-isobaric ensemble}, J. Phys.
  A. Math. Gen. 39~(19) (2006) 5629--5651.
\newblock \href {https://doi.org/10.1088/0305-4470/39/19/S18}
  {\path{doi:10.1088/0305-4470/39/19/S18}}.

\bibitem{thompson2009general}
A.~P. Thompson, S.~J. Plimpton, W.~Mattson, General formulation of pressure and
  stress tensor for arbitrary many-body interaction potentials under periodic
  boundary conditions, The Journal of Chemical Physics 131~(15) (2009) 154107.
\newblock \href {https://doi.org/10.1063/1.3245303}
  {\path{doi:10.1063/1.3245303}}.

\bibitem{hastie2009elements}
T.~Hastie, R.~Tibshirani, J.~Friedman, The Elements of Statistical Learning:
  Data Mining, Inference, and Prediction, Springer, 2009.

\bibitem{scipy}
{SciPy Community}, Scipy documentation,
  \url{https://docs.scipy.org/doc/scipy/reference/optimize.html}, Last accessed
  on 01 Feb 2024 (2023).

\bibitem{nelder1965simplex}
J.~A. Nelder, R.~Mead, A simplex method for function minimization, The Computer
  Journal 7~(4) (1965) 308--313.

\bibitem{Garg1995}
A.~Garg, {Escape-field distribution for escape from a metastable potential well
  subject to a steadily increasing bias field.}, Phys. Rev. B 51~(21) (1995)
  15592--15595.

\bibitem{Dudko2003}
O.~K. Dudko, a.~E. Filippov, J.~Klafter, M.~Urbakh, {Beyond the conventional
  description of dynamic force spectroscopy of adhesion bonds.}, Proc. Natl.
  Acad. Sci. U. S. A. 100~(20) (2003) 11378--81.
\newblock \href {https://doi.org/10.1073/pnas.1534554100}
  {\path{doi:10.1073/pnas.1534554100}}.

\bibitem{Friddle2008a}
R.~Friddle, {Unified Model of Dynamic Forced Barrier Crossing in Single
  Molecules}, Phys. Rev. Lett. 100~(13) (2008) 138302.
\newblock \href {https://doi.org/10.1103/PhysRevLett.100.138302}
  {\path{doi:10.1103/PhysRevLett.100.138302}}.

\bibitem{AbramSteg}
M.~Abramowitz, I.~Stegun, {Handbook of Mathematical functions}, Dover, New
  York, 1972.

\end{thebibliography}

\end{document}